\documentclass[twocolumn,arydshln]{aastex631}

\usepackage{amsmath}
\usepackage{url}
\usepackage{multirow}
\usepackage{color}
\usepackage{float}

\shorttitle{W50 East in X-rays}
\shortauthors{Safi-Harb et al.}

\newcommand\nustar{{\it NuSTAR\/}}

\begin{document}

\title{
Hard X-ray emission from the eastern jet of SS~433 powering the W50 `Manatee' nebula:
Evidence for particle re-acceleration}

\correspondingauthor{Samar Safi-Harb}
\email{samar.safi-harb@umanitoba.ca}

\author[0000-0001-6189-7665]{S. Safi-Harb}
\affiliation{University of Manitoba, 
Department of Physics \& Astronomy,
Winnipeg, MB R3T 2N2, Canada}

\author{B. Mac Intyre}
\affil{University of Manitoba, 
Department of Physics \& Astronomy,
Winnipeg, MB R3T 2N2, Canada}

\author{S. Zhang}
\affil{Bard College Physics Program, 30 Campus Road, Annandale-On-Hudson, NY 12504, USA}

\author{I. Pope}
\affil{Columbia Astrophysics Laboratory, 550 West 120th Street, New York, NY 10027, USA}

\author{S. Zhang}
\affil{Columbia Astrophysics Laboratory, 550 West 120th Street, New York, NY 10027, USA}

\author{N. Saffold}
\affil{Fermi National Accelerator Laboratory, PO Box 500, Batavia IL, 60510, USA}

\author{K. Mori}
\affil{Columbia Astrophysics Laboratory, 550 West 120th Street, New York, NY 10027, USA}

\author{E.~V. Gotthelf}
\affil{Columbia Astrophysics Laboratory, 550 West 120th Street, New York, NY 10027, USA}

\author{F. Aharonian}
\affil{Dublin Institute for Advanced Studies, 31 Fitzwilliam Place, Dublin, Ireland}
\affil{Max-Planck-Institut for Nuclear Physics, P.O. Box 103980, 69029 Heidelberg, Germany}

\author{M. Band}
\affil{University of Manitoba, 
Department of Physics \& Astronomy,
Winnipeg, MB R3T 2N2, Canada}

\author{C. Braun}
\affil{University of Manitoba, 
Department of Physics \& Astronomy,
Winnipeg, MB R3T 2N2, Canada}

\author{K. Fang}
\affil{Department of Physics, Wisconsin IceCube Particle Astrophysics Center, University of Wisconsin, Madison, WI, 53706, USA}

\author{C. Hailey}
\affil{Columbia Astrophysics Laboratory, 550 West 120th Street, New York, NY 10027, USA}

\author{M. Nynka}
\affil{Kavli Institute For Astrophysics and Space Research, Massachusetts Institute of Technology, Cambridge, MA, USA}

\author{C.~D. Rho}
\affil{University of Seoul, Seoul, Rep. of Korea}

\begin{abstract}
We present a broadband X-ray study of W50 (`the Manatee nebula'), the complex region powered by the microquasar SS~433, that provides a test-bed for several important astrophysical processes. The W50 nebula, a Galactic PeVatron candidate, is classified as a supernova remnant but has an unusual double-lobed morphology likely associated with the jets from SS~433. Using  NuSTAR, XMM-Newton, and Chandra observations of the inner eastern lobe of W50, we have detected hard non-thermal X-ray emission up to $\sim$30~keV, originating from a few-arcminute size knotty region (`Head') located $\lesssim$ 18$^{\prime}$ (29~pc for a distance of 5.5~kpc) east of SS~433, and constrain its photon index to 1.58$\pm$0.05 (0.5--30 keV band).  The index gradually steepens eastward out to the radio `ear' where thermal soft X-ray emission with a temperature $kT$$\sim$0.2~keV dominates. The hard X-ray knots mark the location of acceleration sites within the jet and require an equipartition magnetic field of the order of $\gtrsim$12$\mu$G. The unusually hard spectral index from the `Head' region challenges classical particle acceleration processes and points to particle injection and re-acceleration in the sub-relativistic SS~433 jet, as seen in blazars and pulsar wind nebulae.
\end{abstract}


\section{Introduction} \label{sec:intro}

\begin{figure*}
    \centering
   \includegraphics[width=\textwidth]{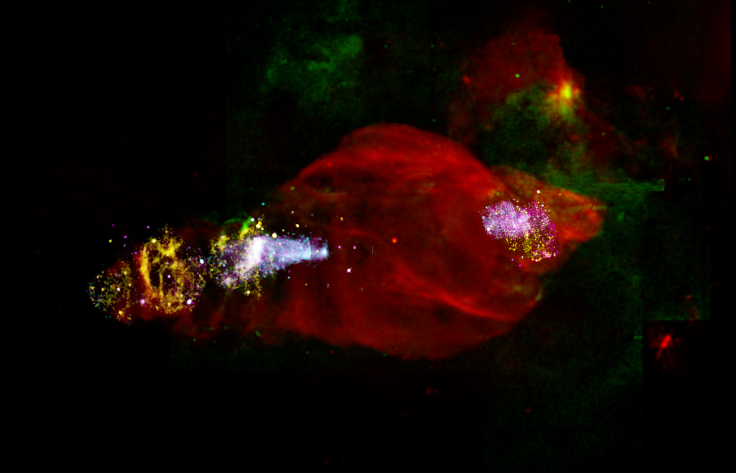}
\caption{
Multi-wavelength image of the W50 nebula. Red: Radio \citep{1998AJ....116.1842D}, Green: Optical \citep{2007MNRAS.381..308B}, Yellow: Soft X-rays (0.5--1 keV), Magenta: Medium energy X-rays (1--2 keV), Cyan: Hard X-ray emission (2--12 keV). The eastern lobe X-ray image highlights the XMM-Newton data presented in this work (with the brightening at the edge of the field of view cropped here to highlight the source emission). The western lobe X-ray image shows only partial Chandra coverage of part of the nebula \citep{2005AdSpR..35.1062M}; additional X-ray observations of this and other regions are currently being carried out and will be presented in future work.}
    \label{fig:beautifulw50}
\end{figure*}

The W50 nebula, catalogued as a Galactic supernova remnant (SNR G39.7--2.0) \citep{1974A&A....32..375V, 1980ApJ...236L..23V},
is best known for its association with the microquasar SS~433, a binary system that displays precessing semi-relativistic ($v$=0.26c) jets, believed to be a  
stellar-mass-sized black hole analogue to Active Galactic Nuclei, AGN {\citep{1984ARA&A..22..507M, 2004ASPRv..12....1F}. With a
linear extent of $\sim$2$^o$ (East-West) $\times$ 1$^o$ (North-South), W50 is one of the largest known SNRs in our Galaxy, over 200 pc $\times$ 100 pc across for the assumed distance $d_{\earth} = 5.5$~kpc \citep{2004ApJ...616L.159B}.
Dubbed the `Manatee nebula' for its radio appearances\footnote{https://www.nrao.edu/pr/2013/w50/}
\citep{1987AJ.....94.1633E, 1998AJ....116.1842D,farnes2017}, the SNR's unique morphology is likely the result of its interaction with the jets from SS~433, as suggested by the location of optical filaments and the elongation of W50 aligned with the axes of the jets' precession cone. This conclusion was also reached using numerical simulations \citep{2011MNRAS.414.2838G}.
The `ears' (eastern and western radio-bright edges) of the nebula are reminiscent of the lobes seen in active galaxies and in the bubbles associated with some Ultra-Luminous X-ray sources. 
Figure~\ref{fig:beautifulw50} shows our multi-wavelength image of this fascinating source.

Regions of the W50 nebula have been sampled within the field-of-view (FoV) of several X-ray missions: ROSAT, ASCA and RXTE in the 1990's
\citep{1997ApJ...483..868S, 1994PASJ...46L.109Y, 1999ApJ...512..784S}
and earlier this century with XMM-Newton and Chandra
\citep{2007A&A...463..611B, 2005AdSpR..35.1062M}.
In 2019, the SRG/eROSITA mission provided an unprecedented view of the whole nebula of W50/SS433 in X-rays.\footnote{Khabibullin and the SRG/eROSITA consortium 2020, https://events.mpe.mpg.de/event/9/contributions/339/}
Overall, the X-ray emission is far from what would be expected from an evolved SNR due to: (a) the morphology of the X-ray lobes filling the gap between SS433 and the radio `ears',  and within the projection of the jets' precessing cone, and, (b) the X-ray spectrum softening away from SS433, along the axes of the precessing jets. Furthermore, while the outermost region of W50, coincident with the eastern radio ear (referred to as region `e3' in \cite{1997ApJ...483..868S}), is dominated by soft thermal X-rays, for the interior (see regions `e1' to `e2' in Figure~\ref{fig:regionsimages}),
hard thermal (kT$\sim$5--12 keV) or non-thermal  (photon index $\Gamma$$\sim$1.4--1.6) emission is inferred \citep{1999ApJ...512..784S}.

\begin{figure*}
    \centering
 \includegraphics[width=\textwidth]{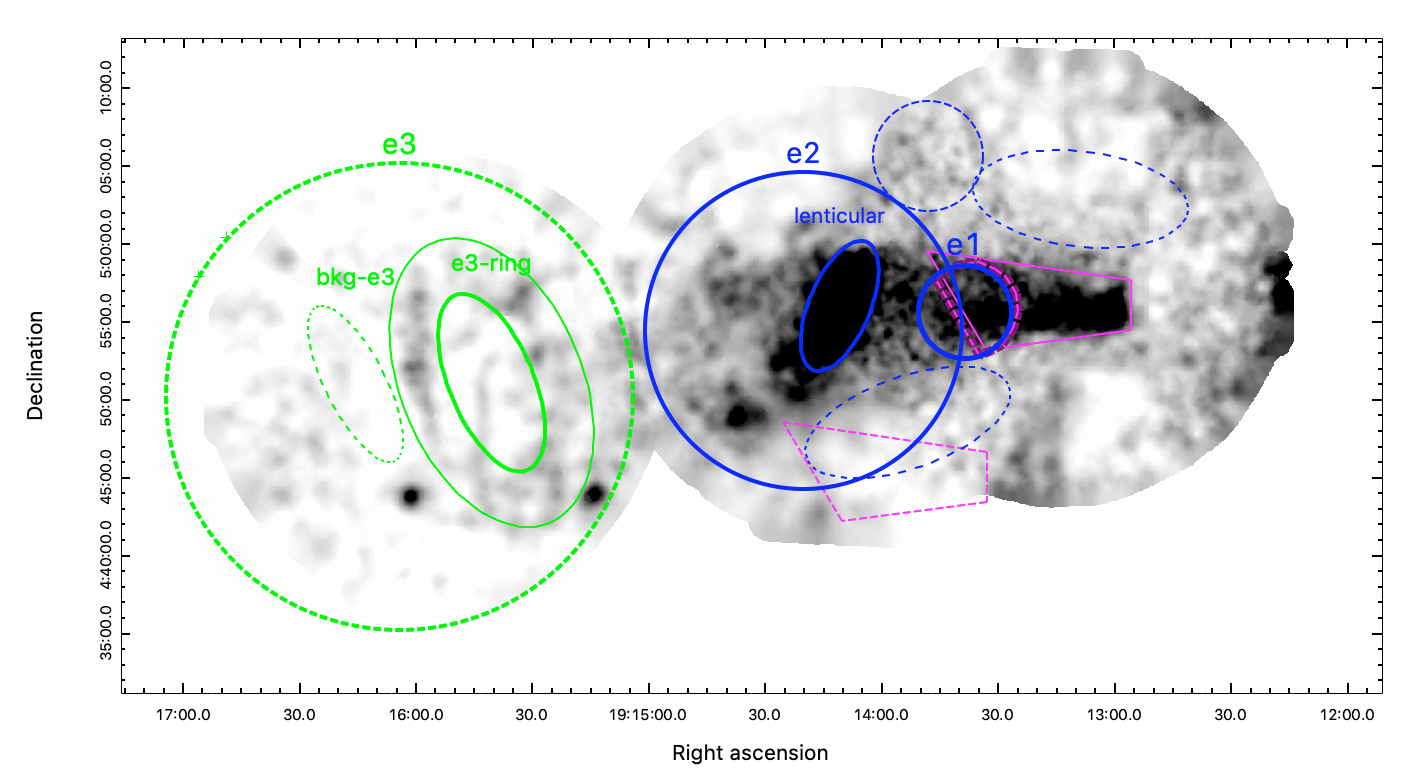}
    \centering
 \includegraphics[width=0.75\textwidth]{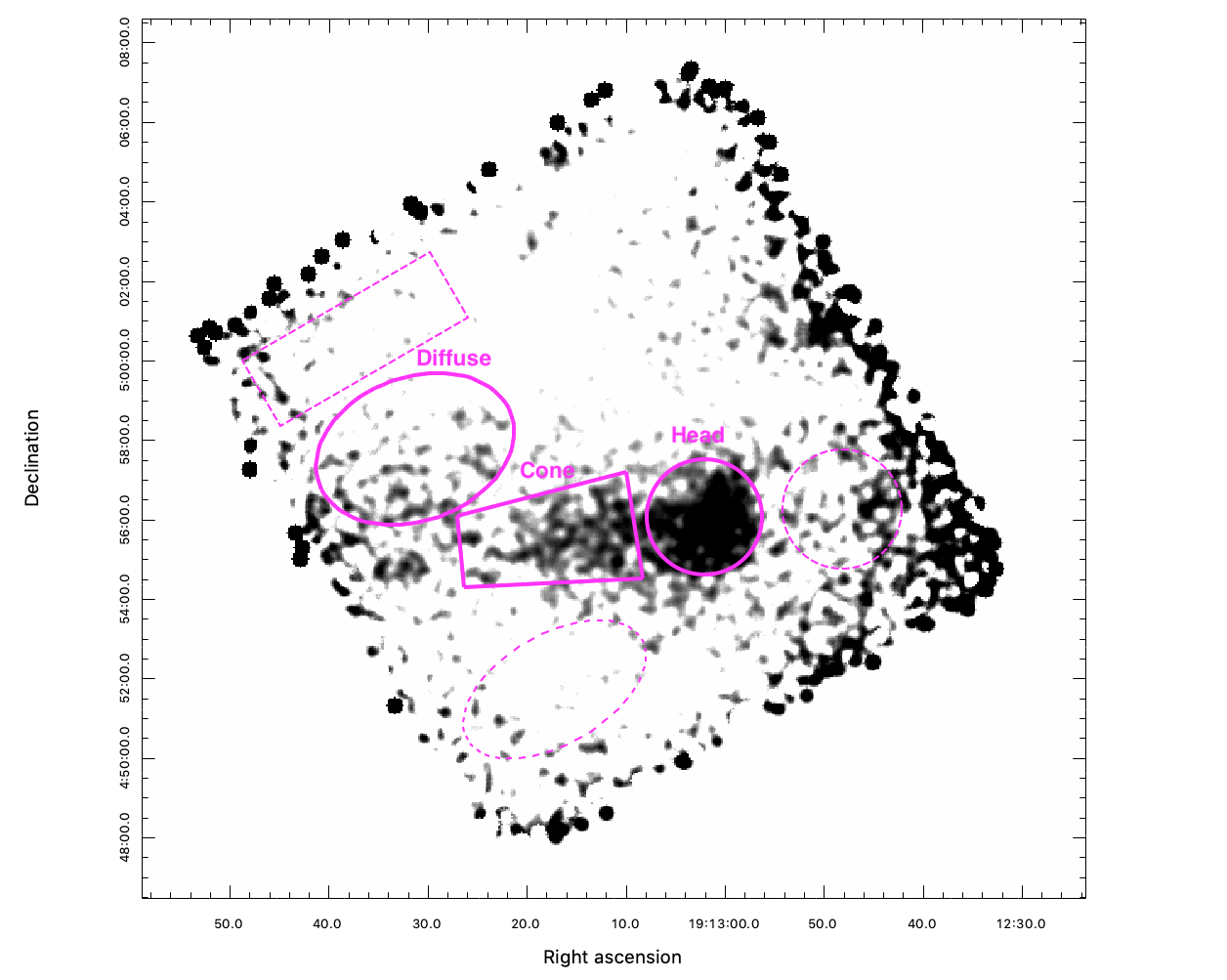}
    \caption{{\bf Top panel:} XMM-Newton 0.3--10 keV mosaic image 
   overlaid with source (solid) and background (dashed) spectral analysis regions. The `e1' and `e2' regions are shown in blue, the `e3' region (which overlaps with the radio `ear') in green. The `lenticular' region is the brightest region within `e2'.
The truncated `Nue1' region in `e1' (purple dashed line) follows the edge of the NuSTAR field-of-view shown below.
{\bf Bottom panel:} The broadband 3--30 keV NuSTAR exposure-corrected and background-subtracted image showing
    the selected regions for the NuSTAR and joint XMM-Newton+NuSTAR spectral analysis.
   In both panels, the bright edges are caused by dividing by the exposure map at the edges of the FoV. 
    The `Head', `Cone' and `Diffuse' spectral analysis regions  fall within the `Full' region shown in the top panel (purple polygon). }
 \label{fig:regionsimages}
\end{figure*}

This hard spectrum implies, 
in the non-thermal interpretation, maximum energies of electrons of $\sim$300--450 TeV responsible for the synchrotron emission,
thus unveiling a new site of cosmic ray acceleration in a jet source: a Galactic PeVatron candidate \citep{1999ApJ...512..784S, 2005A&A...439..635A}.
This motivated the search for high-energy, gamma-ray emission from the source using the Major Atmospheric Gamma Imaging Cherenkov telescopes (MAGIC) and the High Energy Spectroscopic System (H.E.S.S.) \citep{2018A&A...612A..14M}, as well as the Very Energetic Radiation Imaging Telescope Array System (VERITAS) \citep{2017ICRC...35..713K}, leading to  upper limits on TeV emission from the system.
The PeVatron scenario has been more recently revived with the discovery of TeV emission with the High Altitude Water Cherenkov (HAWC) Observatory  \citep{2018Natur.562...82A} whose peak lies close to the `e1' region.
Furthermore, a recent joint Fermi-LAT and HAWC analysis (\cite{2020ApJ...889L...5F} and references therein) finds common emission sites of GeV-to-TeV $\gamma$-rays inside the eastern and western lobes of SS~433.
Despite the surge of gamma-ray studies and multi-wavelength SED modelling of this system,
the origin of the hard(est) X-ray emission, its extent/compactness, and steepness of its spectrum
remain unknown.

In this paper, we present the first joint NuSTAR and XMM-Newton study of the inner eastern lobe of W50 to 
probe the properties of the hard X-ray emission and thus address the particle acceleration process(es) 
powering this Galactic PeVatron candidate. Combining new and  archival XMM-Newton data and Chandra, we also present the first XMM-Newton coverage of the full eastern jet extending from SS~433 out to the radio `ear', as well as the first Chandra coverage of the innermost X-ray emitting region.

\section{Observations}

Three main X-ray emitting regions in the eastern lobe of W50 are defined in \citet{1997ApJ...483..868S} and labelled `e1', `e2' and `e3', centered at 24$^{\prime}$, 35$^{\prime}$, and 60$^{\prime}$ east of SS 433, respectively. For $d_\earth = 5.5$~kpc, these correspond to
 a projected distance from  SS~433 of 38~pc, 56~pc and 96~pc, respectively.
In the following, we describe new NuSTAR and XMM-Newton observations targeting region `e1' along with an archival XMM-Newton observation of  `e2' and `e3' \citep{2007A&A...463..611B}. A Chandra observation of SS~433 \citep{2002Sci...297.1673M}  serendipitously covered a portion of `e1'. Summary images that outline these regions is shown in Figure~\ref{fig:regionsimages}.
A  list of the X-ray observations used in this study is presented in Table~\ref{tab:observationslog}.

\subsection{NuSTAR}

NuSTAR observed the inner eastern lobe region, targeting region `e1',  on 2019 December 1.
NuSTAR consists of two co-aligned X-ray telescopes coupled with focal plane detector modules FPMA and
FPMB \citep{2013ApJ...770..103H}. These are sensitive to X-rays in the 3$-$79~keV band, with a characteristic spectral resolution of 400~eV FWHM at 10~keV. The multi-nested foil mirrors provide $18^{\prime\prime}$ FWHM ($58^{\prime\prime}$ HPD) imaging resolution over a $12\farcm2\times 12\farcm2$ field-of-view. NuSTAR data were processed and analyzed using {\tt FTOOLS} (HEAsoft 6.28)  with NuSTAR Calibration Database (CALDB) files of 2021 January 05. The resulting dataset provides a total of 109~ks,
with 32~ks on a nearby off-source background field.  For all subsequent analysis we merged data from the two FPM detectors. Images and exposure maps were generated following the standard procedure as outlined in \cite{2014ApJ...789...72N}, using FTOOLS extractor and {\tt nuexpomap}. The X-ray background for the source region, which is largely dominated by stray-light photons, is estimated from the scaled off-source observation. The mirror 
vignetting is applied to the exposure correction.

\subsection{XMM-Newton}

XMM-Newton observed regions in the eastern lobe covering `e1' and part of `e2'
on 2020 March 24
for 69~ks.
The European Photon Imaging Camera (EPIC) on-board XMM-Newton
consists of three detectors, the EPIC pn detector \citep{2001A&A...365L..18S}
and EPIC MOS1 and MOS2 \citep{2001A&A...365L..27T}.  
The on-axis point spread
function has a FWHM of $\sim$$12\farcs5$ and $\sim$$4\farcs3$ at
1.5~keV, for the pn and MOS, respectively. The EPIC detectors are
sensitive to X-rays in the 0.15$-$12~keV range with moderate energy
resolution of $E/\Delta E({\rm pn}) \sim $20$-$50.
Data were reduced and analyzed using the Standard Analysis Software
(SAS) v.19.1.0 with the most up-to-date calibration files. The source and background regions were chosen as described in detail in Section~\ref{sec:specana} and extracted using {\tt evselect}. The response matrix file (rmf) and ancillary response file (arf) were generated for extended emission using the SAS commands {\tt{rmfgen}} and {\tt{arfgen}}.
After filtering out background flares we obtained an effective exposure of 31/29/23~ks,
for the MOS1/MOS2/pn data, respectively.
We supplement our study with the archival XMM-Newton observation, carried out in 2004, covering regions `e2' to `e3'. After cleaning the data, the effective exposure time is 26.7~ks for each  MOS1 and MOS2 and 21.4~ks for pn.

\subsection{Chandra}
The innermost portion of the eastern lobe was serendipitously covered during a Chandra ACIS-S\footnote{https://cxc.harvard.edu/proposer/POG/html/chap6.html}
observation obtained on 2000 June 27
and aimed at SS~433. We used this observation to explore the morphology and verify the spectral index of the region overlapping `e1'. We reduce the data using CIAO version 4.12 \citep{2006SPIE.6270E..1VF}
and the calibration files in the CALDB database (version 4.9.1). The event files were reprocessed from level 1 using standard procedures to remove pixel randomization and to correct for CCD charge transfer efficiencies.
Response files were created using {\tt specextract}, and the background was selected from a nearby source-free region south of the source and located on the same chip. The resulting effective exposure time for this observation is 9.7~ks.

\begin{figure*}
    \centering
\includegraphics[width=0.9\textwidth]{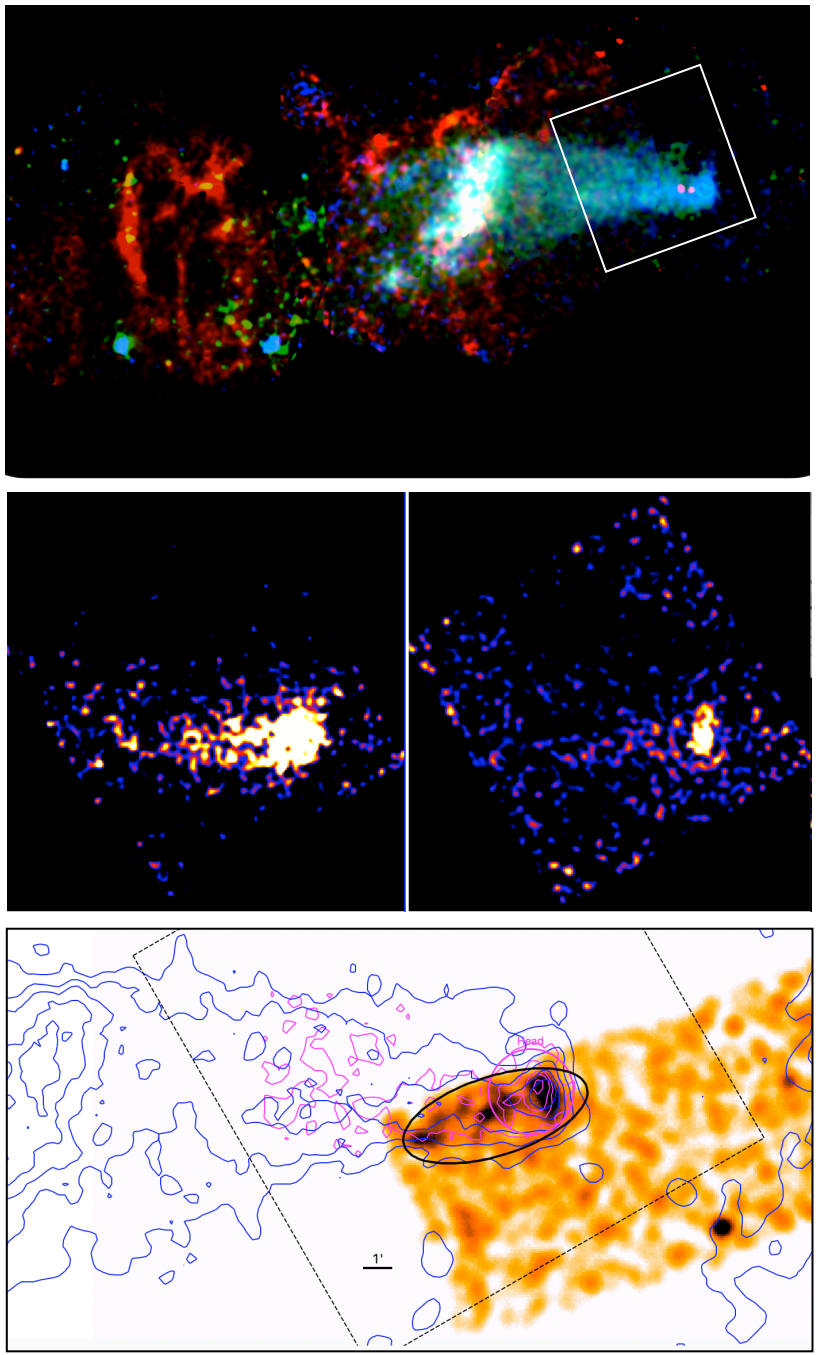}
 \vspace{-2.0cm}
    { \caption{{\bf Top panel}: An energy sliced mosaic image of W50 
     generated by merging exposure-corrected and smoothed new and archival XMM-Newton data. 
     The color scheme correspond to the 0.5--1.0~keV (red), 1.0--2.0~keV (green), and 2--10~keV (blue). 
     The white box outlines the NuSTAR FoV shown in the next panel (containing region `e1'). 
     {\bf Middle panel}: Combined NuSTAR FPMA and FPMB smoothed images in the 3--10~keV (left) and 10--20~keV (right) band,  exposure and vignetting corrected. 
     {\bf Bottom panel}: Serendipitous partial Chandra coverage of the inner part of the eastern lobe. Contours are NuSTAR (magenta) and XMM-Newton (blue). The dashed box is the NuSTAR FoV and the ellipse (black) shows the Chandra spectral extraction region that 
     partially overlaps the `Head' region.
     The image is bounded by SS~433 on the right (west) and the `lenticular' region on the left (east).
     }}
    \label{fig:energyimages}
\end{figure*}

\section{Analysis Results}
     
\subsection{Imaging Analysis}\label{sec:imaging}
Figure~3 (top) shows an energy resolved RGB color image of the eastern lobe  that combines the new and archival XMM-Newton observations.
Red, green and blue correspond to the soft (0.5--1.0 keV), medium (1.0--2.0 keV) and hard (2--10 keV) X-ray bands, respectively.
The colors clearly show that (a) the innermost region of W50 is the hardest and starts at a distance closer to SS~433 than `e1',
(2) there is a mix of hard and soft X-ray emission in `e2', the brightest X-ray emitting region in the eastern lobe (lenticular, white region) where the extension of the radio shell to the North roughly intersects with the axis of the SS433 jet (see 
Section~\ref{sec:discussion} below and Figure~\ref{fig:multiw-ima}), and (3) the 
apparent jet-ISM termination region (coincident with the radio `ear') is dominated by soft X-ray emission with a ring-like structure (in red) that mimics the helical structure seen from the precessing SS~433 jets on much smaller scales \citep{2002Sci...297.1673M}.

Figure~3 (middle)
presents the merged NuSTAR FPMA and FPMB exposure-corrected, background-subtracted image of the `e1' region in the 3--10~keV and 10--20~keV energy bands.  
The X-ray emission is seen to peak in the `Head' region in both the soft and hard X-ray bands, and extending into the `e1' region where the emission becomes fainter and more diffuse.
In Figure~3 (bottom), we show the Chandra image 
with the XMM-Newton and NuSTAR contours overlaid. The `Head' emission fell on the outer S1 chip of ACIS-S and is close to its edge. 

In what follows, we examine the spatial profile of the region around the `Head' to address its extent.

\subsubsection{Spatial Profile with NuSTAR and XMM-Newton}
In order to examine the morphology and hardness ratio throughout the eastern jet, we created linear profiles along the jet for both the soft and hard bands. The orientation and dimensions of the region used to form the linear profiles are shown 
in Figure~\ref{fig:xmmnustarprofiles}.
The profiles were primarily selected in order to study the emission} around the `Head' location and were created by plotting the summed y-axis image values for each point along the x-axis.

\begin{figure}
\begin{center} 
    \begin{minipage}{0.9\columnwidth}
    \centering
     \includegraphics[width=1.0\columnwidth]{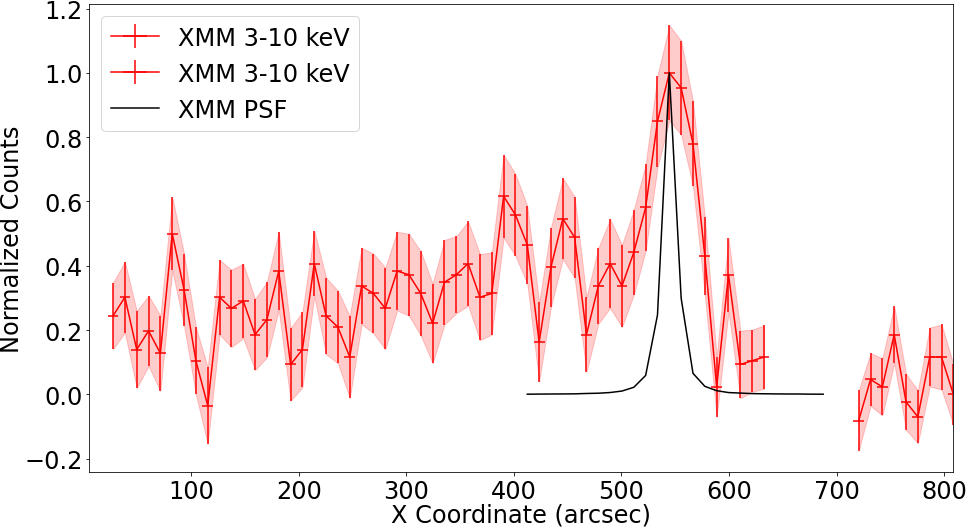}   
    \end{minipage}
    \begin{minipage}{0.9\columnwidth}
    \centering
    \includegraphics[width=\columnwidth]{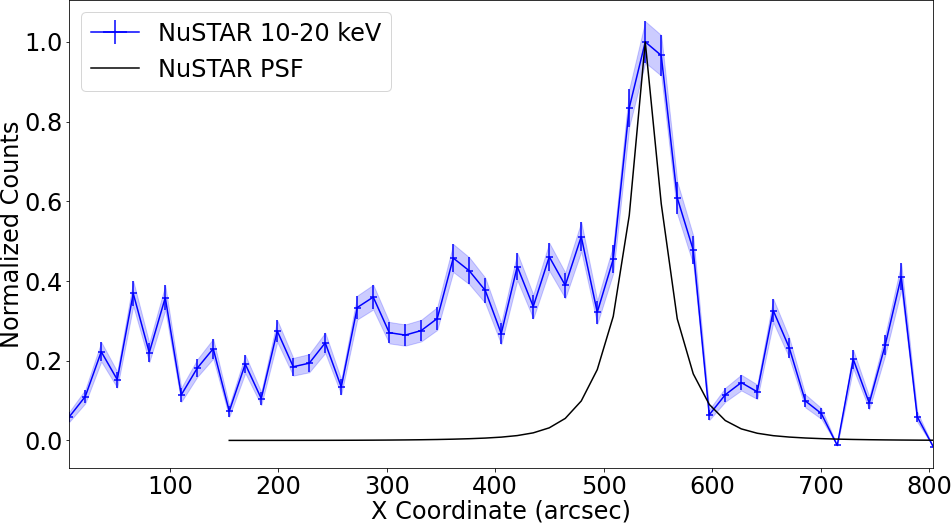}
    \end{minipage}
   \caption{ ({\bf Top}) Background-subtracted spatial profile along the jet for the 3--10~keV XMM image compared to XMM-Newton's PSF. The XMM-Newton image used was binned to 1.1" per pixel in the pipeline. The XMM PSF for 6 keV and 3 arcmin away from on-axis position was used. Binning was set to 10 pixels. ({\bf Bottom}) Background-subtracted spatial profile along the jet for the 10--20~keV NuSTAR image compared to NuSTAR's PSF. Binning was set to 6 pixels.}
   \end{center}
     \label{fig:xmmnustarprofiles}
\end{figure}

We first sought to compare linear profiles of the jet around the `Head' region to the PSF of the telescope used for the observation. For the soft band, we used the XMM-Newton observation because of its superior angular resolution compared to NuSTAR. For this analysis, we only used the MOS1 image and initially binned it to 1.1$^{\prime}$ per pixel.
The profile was then binned to 10 pixels in order to smooth it without losing considerable structural information.
For the background subtraction, we chose a region close to the source region and subsequently subtracted the background region profile from the source profile.
The NuSTAR observation was used to create the hard band linear profile. With a pixel size of 2.46$^{\prime\prime}$, we chose to bin the \nustar\  profile to 6 pixels per bin. The x-axis values of the profiles were then scaled by their respective observation's pixel size in arcseconds in order to make it easier to compare the XMM-Newton and NuSTAR profiles. 
The NuSTAR background subtraction was done by subtracting the processed  off source image from the processed on source image.

For comparison to the XMM MOS1 3--10 keV profile, we use the MOS1 6 keV, 3$^{\prime}$ off-axis PSF. We compare the NuSTAR 10--20 keV profile to a NuSTAR 12--20 keV effective PSF. We normalized both the linear profiles and PSFs to unity. The XMM-Newton MOS1 and NuSTAR PSFs were then centered upon the highest value of their respective profiles. The XMM 3--10 keV profile and corresponding PSF are shown in Figure~\ref{fig:xmmnustarprofiles} (top),
whereas the NuSTAR 10--20 keV and corresponding PSF are shown in Figure~\ref{fig:xmmnustarprofiles} (bottom).
We note that the gaps seen in the XMM-Newton linear profile demarcate the MOS1 chip gap.
These plots show that the `Head' region is extended compared to the nominal instrument's PSF.

\begin{figure}
\begin{center} 
 \includegraphics[width=0.45\textwidth]{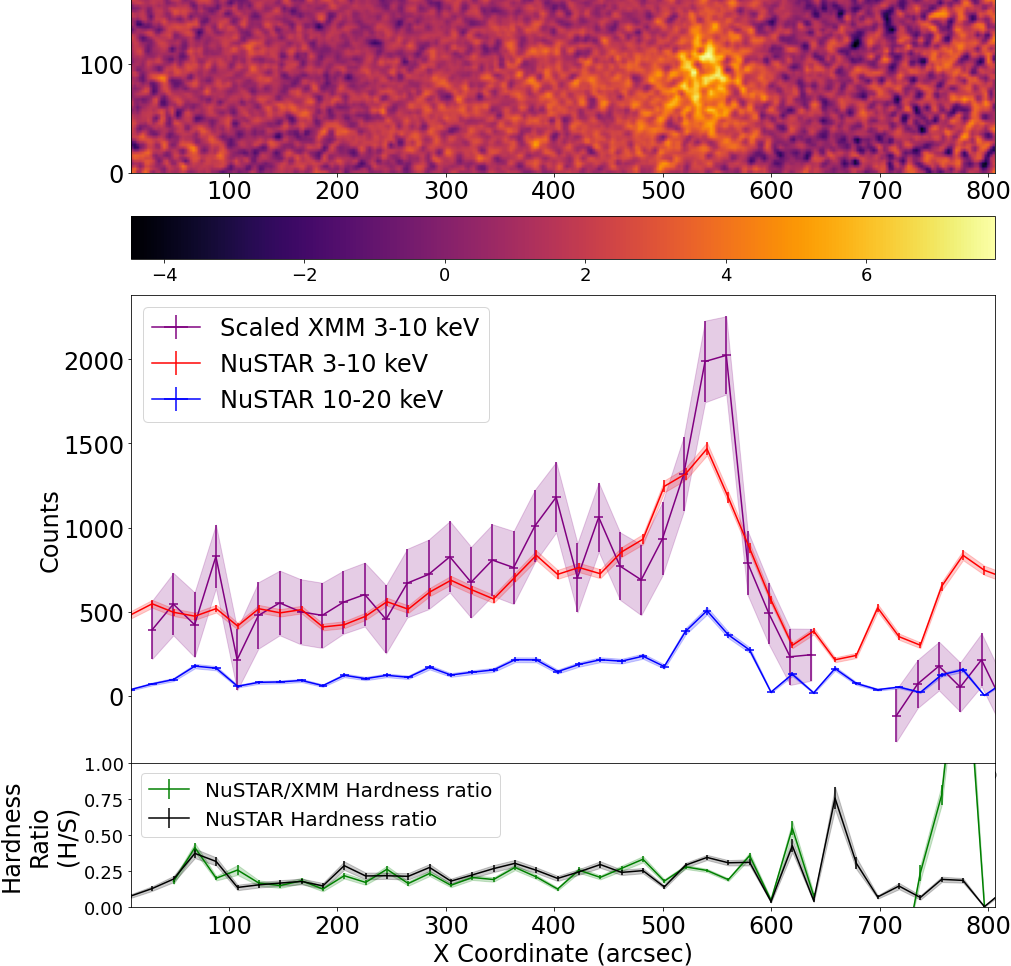}
  \caption{The scaled XMM-Newton and NuSTAR profiles around the `Head' region (top) and the hardness ratio (bottom) which diverge from each other towards SS~433 (to the right of the `Head' region), where we expect the greatest contribution from ghost rays.}
  \label{fig:nustarhardness}
\end{center}
\end{figure}

In order to measure the hardness ratio along the jet, we must compare the soft and hard band linear profiles. 
The NuSTAR observation is however affected by ghost rays near the `Head' region. The ghost ray contribution is more significant in the soft band, so we attempt to avoid these ghost rays by also comparing the XMM-Newton soft band profile and the NuSTAR hard band profile. We bin both the NuSTAR 3--10 keV and 10--20 keV linear profiles to 8 pixels per bin. For the XMM 3--10 keV profile, we initially bin the MOS1 image to 2.45$^{\prime\prime}$ per pixel, close to NuSTAR's 2.46$^{\prime\prime}$ pixel size. We then bin the XMM 3--10 keV profile to 8 pixels per bin. 
In order to make the XMM-Newton soft band profile comparable to the NuSTAR profiles, we scale the XMM soft band profile so that the average value of the leftmost 15 bins is normalized to the corresponding extrapolated values of the NuSTAR soft band profile. By choosing the leftmost 15 bins, we avoid the ghost rays near the `Head' region of the jet. 
We note that for this profile comparison, the background subtraction was done before scaling the XMM-Newton profile to the NuSTAR soft band profile.
In each case, we calculate the hardness ration as $H/S$,
where $H$ is the number of counts from the hard band profile and $S$ is the number of counts from the soft band profile.
The various profiles and hardness ratio plots are shown in 
Figure~\ref{fig:nustarhardness}. The NuSTAR-only and NuSTAR-XMM hardness ratio plots are similar, diverging from each other the most in the region to the right of the `Head', where we expect the greatest contribution from ghost rays.
Ghost rays are further discussed in \S\ref{sec:nustarbkg}.

\subsubsection{Chandra}

\begin{figure}
   \centering
 \includegraphics[width=1.05\columnwidth]{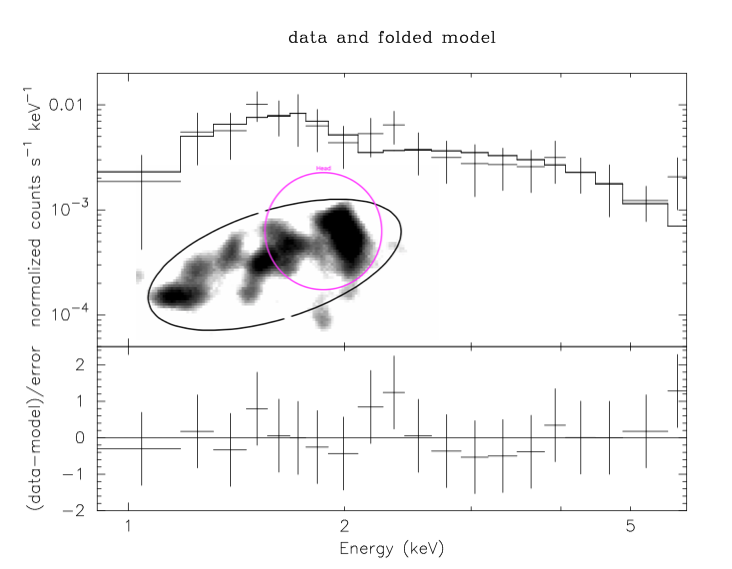}
\begin{scriptsize}    \caption{
Chandra image (Inset) and spectrum of the inner eastern lobe overlapping with the `Head' region (purple circle overlaid on the image).
The image is shown on a linear scale (from 0.5 to 5 counts per pixel) and is smoothed with a Gaussian with $\sigma$ of 7 pixels.
The overlaid black ellipse shows the region used for extraction of the Chandra ACIS spectrum fitted
with an absorbed power-law model with a photon index $\Gamma$$\sim$1.5 
(see \S\ref{sec:chandraspectroscopy}).}
    \end{scriptsize}
   \label{fig:chandraimagespec}
\end{figure}

In order to further verify the extent of the hard X-ray emitting region and spatially resolve it, we examine
the archival Chandra observation of SS~433.
As illustrated in 
Fig.~6 (inset) and Fig.~3 (bottom),
this emission can be resolved into $\sim$1$^{\prime}$-scale knotty structures,
which translate to a scale of $\sim$1.6~pc at a distance $d_{\earth} = 5.5$~kpc.
This Chandra dataset is however limited by poor statistics (see \S\ref{sec:chandraspectroscopy}). A more detailed analysis requires a deeper and targeted Chandra observation of this region.

\subsection{Spectral Analysis}

\label{sec:specana}
Guided by previous X-ray observations of the W50 system (see \S\ref{sec:intro}) we considered both non-thermal (power-law and broken power-law) models and thermal
models for our spectral fits. 
Spectral analysis was accomplished using the X-ray spectral fitting package XSPEC v12.11.1~\citep{1996ASPC..101...17A}\footnote{http://xspec.gsfc.nasa.gov}
with the {\tt tbabs model} to describe photoelectric absorption by the interstellar medium,  using the \cite{2000ApJ...542..914W} abundances. 
In general, this yields column densities, $N_{\rm H}$, that are higher than previously published with earlier abundance models. 

The XMM-Newton, NuSTAR, and Chandra data were fitted in the 0.5--10 keV,  3--30 keV, and 0.5--8.0 keV band respectively. 
The XMM-Newton spectra were grouped with a minimum of 30/50 counts per bin for MOS/pn data and 10 counts per bin for Chandra. For NuSTAR, the background subtracted spectra are grouped for at least $3\sigma$ detection level per data bin.
Since NuSTAR is not sensitive to the soft X-ray band,  we fixed the column density to the value derived from XMM-Newton. The Chandra spectrum was fitted independently given its coverage of an overlapping, but not identical, region in the innermost eastern lobe (Fig.~2). 
For all joint fits, a constant factor is used to account for calibration differences between instruments. 
Quoted errors are at the 90\% confidence level (CL), and our fitting procedure used the chi-square ($\chi^2$) statistic, unless otherwise noted.

\subsubsection{NuSTAR and XMM-Newton Spatially Resolved Spectroscopy}\label{sec:spresspec}
Careful background subtraction is needed for the spectral analysis of the W50 lobe regions, given the nearness of W50 to the Galactic plane and its bright engine SS~433. This results in a complicated NuSTAR background
due to stray light, ghost rays, instrumental, and cosmic background components. Our NuSTAR background analysis is detailed in Appendix~\S\ref{sec:nustarbkg}.
For the XMM-Newton data, background selections consistent between archival and new observations were made for verification. Additionally,  background selections identical to those used for the NUSTAR spectral analysis were used as an additional layer of consistency check.
Over 13 background region selections, including from a nearby off-source observation, were tested before settling on the background selection shown in 
Figure~\ref{fig:regionsimages}.
Overall, we opted for background regions that are closest to the source, source-free, and overlapping in the XMM-Newton and NuSTAR coverage given we are performing a joint spectroscopic study of the same regions selected from both telescopes.

In the following we concentrate on  the hard X-ray emission from W50-East, the main focus of this work, guided primarily by the NuSTAR imaging analysis. 
We divided the NuSTAR spatial coverage into the three sub-regions shown in Figure~\ref{fig:regionsimages},
the `Head', `Cone' and `Diffuse'.
We also extracted a spectrum from a polygon (`Full') region to include all the above three sub-regions.
Lastly, we also examine region `Nue1', since it corresponds to the (significant) portion of the well studied `e1' region bounded by the edge of NuSTAR's FoV and significantly overlaps with the `Diffuse' region.
For each of these regions, we fitted the XMM-Newton spectra (0.5--10 keV) and the NUSTAR spectra (3--30 keV) both independently and jointly to an absorbed power-law model ({\tt const*tbabs*powerlaw} in XSPEC). For the independent NuSTAR fits, we fixed the column density to the XMM-Newton value.
All spectral results are presented in Table~\ref{tab:pow0.5-30}.

\begin{figure*}
    \centering
\includegraphics[width=\textwidth]{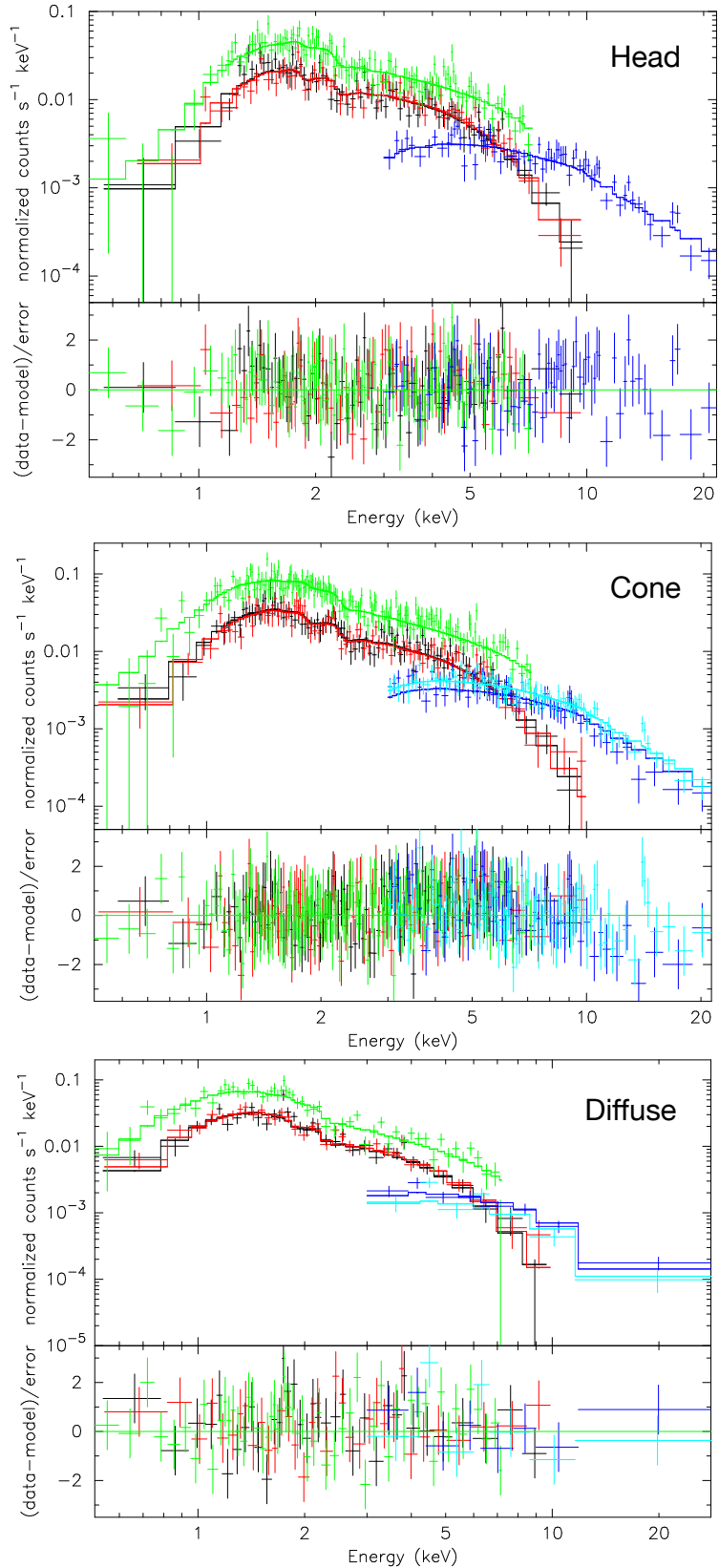}
\caption{The joint XMM-Newton and NuSTAR spectra for the `Head' (top), `Cone' (Middle) and `Diffuse' (bottom)
NuSTAR regions, fitted with the absorbed power-law model as described in the text.
Black, red and green spectra correspond to MOS1, MOS2 and pn, respectively; blue and cyan to the NuSTAR FPMA and FPMB spectra.
The regions are shown on Figure~\ref{fig:regionsimages} and the spectral parameters for these regions are summarized in Table~\ref{tab:pow0.5-30}.}
\label{fig:xmmnustarspec}
\end{figure*}

First, we find that $N_{\rm H}$ varies significantly across the inner eastern lobe, from
$1.77_{-0.17}^{+0.18}$~$\times$~10$^{22}$~cm$^{-2}$ in the `Head' region, to $1.18_{-0.10}^{+0.11}$~$\times$~10$^{22}$~cm$^{-2}$ in the `Cone' region, to $0.76_{-0.10}^{+0.11}$~$\times$~10$^{22}$~cm$^{-2}$ in the `Diffuse' region. 
Going further east, the column density in the `lenticular' and `e3-ring' regions are comparable to that of the Diffuse region (at $\sim$0.7$\times$~10$^{22}$~cm$^{-2}$).
The spectral parameters shown in Table~\ref{tab:pow0.5-30} correspond to the column density, $N_{\rm H}$, frozen to its best fit value determined from the XMM-Newton fits. We find that all regions are 
 adequately described by a power-law model with a photon index of 1.58$\pm$0.05, 1.76$\pm$0.04, and 1.77$\pm$0.06 (0.5--30 keV) for the `Head', `Cone' and `Diffuse' regions, respectively.
The corresponding spectra and fitted model are shown in Figure~\ref{fig:xmmnustarspec}.
When we allow $N_H$ to vary, the errorbars increase somewhat (as expected) but our results do not change (within error). These fits yield a photon index of 1.6$\pm$0.1, 1.86$\pm$0.08, and 1.81$\pm$0.13 (0.5--30 keV) for the `Head', `Cone' and `Diffuse' regions, respectively.
Taking into account the spectral results for the `lenticular' region (Table~\ref{tab:pow0.5-30}), we conclude that there is a gradual spectral softening eastward from SS~433 with the hardest emission at the `Head' region,
the onset of X-ray emission in the eastern lobe.

The index for the `Head' region is consistent between the two instruments (at $\Gamma$$\sim$1.6), whereas for the other regions, NuSTAR gave a slightly steeper index than XMM-Newton. For the larger scale region `Full' which encompasses the source emission across the NuSTAR field, the discrepancy is primarily
caused by the mixture of multiple components with different spectral indices, as demonstrated by our spatially resolved spectroscopy.
For the fainter `Cone' and `Diffuse' regions, 
we fitted a broken power-law model 
to test for any steepening in the NuSTAR hard band above 10~keV. 
We find that this model improves on the power-law model fit for the `Cone' region, but not for the `Diffuse' nor the `Head' regions.
For the `Cone', we obtain $\Gamma_1$~=~1.64$^{+0.04}_{-0.07}$, $\Gamma_2$~=~2.1$\pm$0.1, and a break energy $E_b$~=5.3$^{+0.4}_{-0.8}$~keV (F-test probability of 1.1$\times$10$^{-10}$).
This is likely attributed to mimicking the spectral evolution (steepening) across the wide `Cone' region extending from the `Head' towards the `lenticular' region whose index steepens to $\sim$2.0. 

Finally, we fitted the NuSTAR selected regions with a thermal bremsstrahlung model and measured temperatures of $kT$ = 18.6~(15.6--22.7)~keV, 10.4~(9.4--11.4)~keV and 6.9~(6.0--8.0)~keV for the `Head', `Cone' and `Diffuse' regions, respectively. The unrealistically high temperatures, lack of emission lines, and no statistical improvement of these fits when compared to the power-law fits make us favour the power-law model over thermal models. 
This conclusion is in agreement with previous studies of the eastern lobe (e.g., \cite{1999ApJ...512..784S} and references therein).

Further away from SS~433 towards the radio ear, 
the power-law photon index of the shocked bright regions (covered by XMM-Newton only)
steepens to $\sim$2 (`lenticular' and 
`e3-ring' regions), with
soft thermal emission becoming evident towards these regions
(see Table~\ref{tab:pow0.5-30} and Fig.~3).
The emission becomes dominated by a thermal component with $kT$$\sim$0.2~keV in the termination shock region (`e3-ring') (see also \cite{2007A&A...463..611B, 1997ApJ...483..868S}). 
For this region, the total 0.5--10 keV flux is 1.2$\times$10$^{-12}$~erg~cm$^{-2}$~s$^{-1}$ (observed) and 1.4$\times$10$^{-11}$~erg~cm$^{-2}$~s$^{-1}$ (unabsorbed). The thermal component constitutes approximately 46\% (absorbed flux of 5.6$\times$10$^{-13}$~erg~cm$^{-2}$~s$^{-1}$) and 93\% (unabsorbed flux of 1.3$\times$10$^{-11}$~erg~cm$^{-2}$~s$^{-1}$) of the total flux.
The properties of the thermal X-ray emission are however poorly constrained given its low-surface brightness and require a deep and targeted XMM-Newton observation.

\subsubsection{Chandra Spatially Resolved Spectroscopy} \label{sec:chandraspectroscopy}

We extract an elliptical region (see Fig.~3 and Fig.~6)
encompassing the `Head' region on ACIS S1 for the source and a nearby region for the background subtraction. This source significantly overlaps with the `Head' and `Cone' regions and its
emission is dominated by that of the `Head' which was not well resolved with NuSTAR.
We obtain 243 background-subtracted source counts
and a spectrum best fitted with a power law model with 
$N_{\rm H}$~=~1.61$^{+0.78}_{-0.62}$~$\times$~10$^{22}$~cm$^{-2}$, $\Gamma$ = 1.48$^{+0.50}_{-0.45}$ (1$\sigma$ error range),
and a flux of 5.7$\times$10$^{-13}$~ergs~cm$^{-2}$~s$^{-1}$ (0.5--10~keV).
Freezing the column density to 1.77$\times$10$^{22}$~cm$^{-2}$, the value obtained from XMM-Newton (which is more sensitive to the soft X-ray band) for the `Head' region, we find $\Gamma$ = 1.57$\pm$0.39 (90\% CL), an index which is consistent with our joint spectroscopy~(\S\ref{sec:spresspec}).
We have also attempted different binning for the Chandra spectrum as well as fitting with the Cash statistic. We find a photon index that is consistent with the above-mentioned value, within error.
For example, binning the spectrum with a minimum of 20 counts per bin, we obtain a photon index  $\Gamma$ = 1.6$\pm$0.4 (90\% CL) for $N_{\rm H}$=1.77$\times$10$^{22}$~cm$^{-2}$. 
This confirms the unusually hard photon index obtained for the `Head' region.

\section{Discussion\label{sec:discussion}}

\begin{figure*}
    \centering
\includegraphics[width=\textwidth]{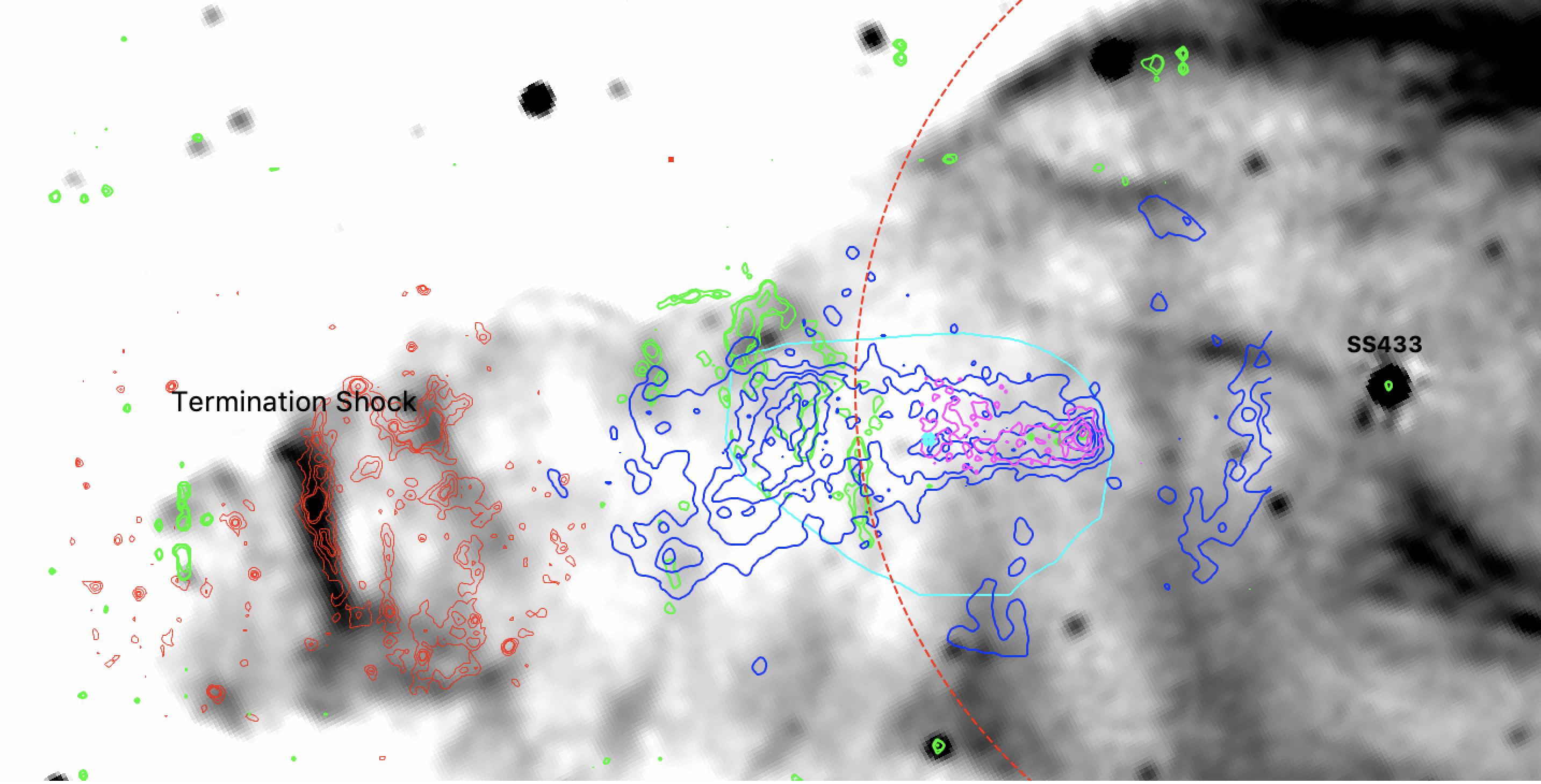}
    \centering
    \caption{Multi-wavelength image of the eastern lobe of W50 bounded by SS~433 to the right and the radio `ear'
    to the left. The radio 1.4~GHz image \citep{1998AJ....116.1842D} is shown with X-ray, optical \citep{2007MNRAS.381..308B} and the HAWC emission \citep{2018Natur.562...82A}.
    Contours shown are XMM-Newton contours of the inner eastern jet covering regions e1--e2 (blue), NuSTAR coverage of the innermost region jet covering e1 (purple), XMM-Newton contours of the e3-ring region (red) overlapping the radio `ear' (or termination shock), optical filaments (green) and the HAWC 4$\sigma$ contour (cyan). We note that
    the blue contour running in the north-south direction in between SS~433 and the head-e1 region delineates the edge of the XMM-Newton field of view. The dashed radio circle denotes the extent of the radio shell seen to the north and south. 
    The inner X-ray lobe is sitting within a radio `hole' (or depression in radio emission), while the optical filaments partially overlap with the projection of the radio shell along the SS~433 jet axis.   }
    \label{fig:multiw-ima}
\end{figure*}

In Fig.~\ref{fig:multiw-ima}, we show the multi-wavelength view of W50-east.
The inner, hard, X-ray lobe is sitting within a radio ‘hole’ (or depression in radio emission), while the
optical filaments partially overlap with the projection of the radio shell along the SS 433 jet axis.
This multi-wavelength view and spatially resolved spectroscopic study supports the picture that the `head' 
represents an acceleration zone and the onset of the eastern lobe X-ray emission; the jet
thermalises first at its interaction with the SNR (`lenticular' region in `e2' which is
just beyond the intersection of the extrapolated radio shell and jet axis), then at its termination shock with the ISM (radio `ear').

Our study allowed us to zoom in on a micro-version of a quasar 
(Fig.~\ref{fig:multiw-ima} and Fig.~\ref{fig:beautifulw50}).
 In radio, W50 mimics FRII galaxies (with prominent lobes resulting from the interaction with the surroundings) but in X-rays, it mimics FRI galaxies dominated by emission from the jets close to the black hole.
However, unlike FRIs, SS~433 is a super-Eddington source, accreting at a super-Eddington rate
of $\sim$10$^{-4}$~$M_{\odot}$~year$^{-1}$ (via Roche-lobe overflow from a massive donor star that is likely an evolved supergiant).
FRIs have much lower Eddington ratios, making SS433 the nearest example of jetted super-Eddington black hole (for a comprehensive review, see \cite{2004ASPRv..12....1F}).
For W50, the jets' large-scale extended emission peaks in X-rays (as opposed to radio) given the high Lorentz factor ($\gamma$ $\gtrsim$ 10$^7$) of the synchrotron emitting electrons at tens of parsecs away from SS~433.
Furthermore, while the nature of X-ray emission in AGN jets is being debated (synchrotron vs inverse Compton)
and is more likely due to the synchrotron process in FRIs \citep{2016MNRAS.455.3526H}, 
the non-thermal X-ray emission in W50 is most likely synchrotron emission.
This is the case because generally, synchrotron X-ray emission has been detected in microquasar jets interacting with the ISM
(see e.g., \cite{2002Sci...298..196C}). Furthermore, for the estimated magnetic field in the eastern lobe (see more below) and hard X-ray photon index, synchrotron emission should dominate the inverse Compton flux. We add that the non-thermal bremsstrahlung process can be also ruled out since the inverse Compton flux should dominate the non-thermal bremsstrahlung flux
for the electron densities inferred in the lobe (see \cite{1999ApJ...512..784S, 1997ApJ...483..868S}).

The SS~433/W50 system also bears resemblance to Ultra Luminous X-ray sources (ULX's), a growing class of X-ray discovered compact objects believed to be associated with black holes (of debated mass) and accreting neutron stars.  ULX's are bright extragalactic point-like sources whose X-ray luminosity ($L_x$ $\gtrsim$ 10$^{39}$~erg~s$^{-1}$) exceeds the Eddington limit for accretion into a stellar-mass black hole radiating isotropically (see \cite{2017ARA&A..55..303K} for a recent review). Interestingly, SS~433 has been suggested to be the nearest Galactic ULX analog \citep{2001IAUS..205..268F}, 
where its radiation can be collimated by the geometry of a funnel and beamed by the motion of the source along the line-of-sight direction (see \cite{2004ASPRv..12....1F} and references therein). However, 
SS~433 is aligned edge-on (or viewed from the side) with respect to Earth \citep{2006MNRAS.370..399B}, and
its core is mostly hidden from view since it is blocked by Compton-thick material present in the plane of the accretion disk.
As a result, SS~433 does not manifest itself a a typical ULX (where the disk wind would be sufficiently optically thin to allow a copious amount of X-ray photons to reach us), nor even as a luminous 
X-ray binary.

Bubble-like large ($\sim$100--500~pc scale) shock-ionized nebulae have been detected, primarily in the radio and optical wavelengths, around some ULX's. Jets have been proposed to be responsible for inflating the observed bubbles (see e.g. \cite{2010Natur.466..209P}). A direct comparison with W50 would then allow us to revisit
the question raised in the 1980's \citep{1980ApJ...238..722B} on whether W50 can be similarly a bubble (or `beam-bag') inflated solely by the SS~433 jets 
(see \S5 in \cite{1997ApJ...483..868S} for a discussion on the beam-bag scenario versus the favoured SNR+jets scenario later verified by numerical simulations \citep{2011MNRAS.414.2838G}).
Two notable examples similar to SS~433 (in terms of the presence of powerful jets with a kinetic power
exceeding 10$^{39}$~erg~s$^{-1}$) are NGC7793--S26 located at a distance of $\sim$3.9~Mpc \citep{2010Natur.466..209P, 2010MNRAS.409..541S} and Holmberg II X--1 located at a distance of $\sim$3.39~Mpc \citep{2014MNRAS.439L...1C, 2015MNRAS.452...24C}.
For S26, a 300~pc diameter jet-inflated-bubble in the galaxy NGC~7793, Chandra resolved two `hot spots', in addition to the core, separated by a projected distance of  15$^{\prime\prime}$ which translates to $\sim$290~pc. A non-thermal (power-law or broken power-law) spectrum has been ruled out for these X-ray emitting hot spots whose spectra are best fitted by a two-component thermal plasma model ($kT_1$$\sim$0.26~keV and $kT_2$$\sim$0.96~keV). Based on their physical scale and X-ray spectrum, we infer that these regions could be the equivalent to the termination shock regions seen in the `ears' of W50 (region `e3' in Fig.~\ref{fig:regionsimages}, see also Fig.~\ref{fig:multiw-ima}). 
The morphology (large radio cocoon and hot thermal X-ray spots) resembles W50, however the total energy content of the W50 bubble is two orders of magnitude lower than that of S26 \citep{2010Natur.466..209P}.
For Holmberg II X-1, the X-ray observations (with Swift and Chandra) show steady, apparently hard spectra, however the X-ray emission is dominated by the emission from the core component and no extended X-ray emission or hot spots have been reported.
Recent attempts to detect X-rays (using Chandra) from Holmberg IX X-1, another large ($\sim$300~pc scale,
distance of 3.8~Mpc) ULX bubble,  
failed to detect any extended X-ray emission down to a luminosity of 2$\times$10$^{36}$~erg~s$^{-1}$ \citep{2019MNRAS.488.4614S}.
Lastly, Chandra observations of the S10 optical nebula in NGC~300 revealed
four discrete X-ray knots aligned in the plane of the sky over a length of $\sim$150~pc \citep{2019MNRAS.482.2389U}. 
Radio observations show an elongated radio
nebula ($\sim$170~pc $\times$ 55~pc in size) with its major axis aligned with the axis connecting the Chandra sources. The X-ray emission from the X-ray knots is thermal ($kT$$\sim$0.6~keV), and believed to arise from the interaction between the engine (microquasar candidate) and the ISM.
These properties are overall strikingly similar to W50, although (unlike W50) there is no detection of non-thermal X-ray emission in NGC~300--S10.

Based on the previously summarized observations, we conclude that while W50 shares many properties with ULX-inflated bubbles,
a sensitive and detailed X-ray study of extragalactic ULX bubbles is currently hampered by their large distances, making their X-ray detection and characterization dominated by the core component (black hole) and/or the relatively bright termination shock regions on the larger scale. The latter appear to be dominated by thermal X-ray emission and is poorly resolved given the large distances (Mpc) in comparison to Galactic distances (kpc). This again makes W50/SS~433 the most accessible ULX-bubble analog for zooming in into the outflows of these energetic sources (from close to the microquasar all the way to the termination shock) and studying their large-scale impact on their surroundings. 
Furthermore, while no non-thermal X-ray emission has been reported to date from the large-scale extragalactic ULX bubbles, deep and high-resolution observations in the hard ($\geq$2~keV) X-ray band, with current and proposed future missions such as AXIS \citep{2019BAAS...51g.107M}, may reveal the equivalent of the non-thermal emission seen in W50's inner lobes on tens of parsecs scales. This emission provides clues on the jet's interaction with the ISM and the debated particle acceleration processes in play.

In this work, we have mapped the broadband X-ray emission of W50-East from SS~433 out to the radio `ear',
 discovered that the hardest non-thermal X-ray emission is from the inner eastern lobe,
and that its peak is resolved  to the `Head' region, 
$\sim$18$^{\prime}$ east of SS~433 ($\sim$29 pc @ $d_{\earth} = 5.5$~kpc).
This emission is characterized by a hard, and rather unusual, spectral index.
A photon index $\Gamma$ of $\sim$1.5 implies a particle index $p$ of $\sim$2 ($\frac{dN}{dE}~\sim~E^{-p}$; $p$=2$\alpha$+1 where $\alpha$=$\Gamma$-1) and therefore an $E^{-2}$ distribution of electrons.
The classical Diffusive Shock Acceleration (DSA) theory predicts a spectrum steeper than our observed X-ray spectrum. This index is in fact more similar to what is observed in pulsar wind nebulae (PWNe) powered by the relativistic winds of neutron stars \citep{2003ApJ...591..361G, 2008AIPC..983..171K} and in some extragalactic jets \citep{2003ASPC..300..151H}. X-ray knots in many AGN jets can also have a photon index $\Gamma$$\sim$1.1--1.6 \citep{2004ApJ...608..698S}.
Furthermore, such energetic electrons should be cooled, thus to have an  $E^{-2}$ type spectrum,  one should have an 
extremely hard differential injection spectrum, harder than $E^{-1}$. A possible explanation within the DSA model could be
that adiabatic losses dominate over synchrotron losses as suggested for blazars to explain their hard spectra \citep{2011ApJ...740...64L}.
Notably, the index here steepens to an $E^{-3}$ (for $\Gamma$$\sim$2) further out as the jet thermalises. 
Therefore it is possible that particles are accelerated in the `Head' region with an $E^{-2}$ spectrum. 
Inside the accelerator, the spectrum is not deformed (due to the adiabatic losses or
energy-independent escape), but electrons, after they leave the source, suffer synchrotron losses which naturally can explain the steepening of the X-ray spectrum in the `Cone' and `Diffuse' regions. 
An alternative acceleration mechanism could be magnetic reconnection or stochastic acceleration which can produce a very hard (e.g. Maxwellian type) acceleration spectrum \citep{2014ApJ...783L..21S, 2015ASSL..407..311L} which after synchrotron cooling becomes $E^{-2}$ (independent of the acceleration spectrum).
We note that the column density is highest in the `Head' region (see Section~\ref{sec:specana}). This could be
evidence of jet entrainment which would contribute to mixing and turbulence impacting the shock acceleration process, e.g. through internal shocks \citep{2013A&A...558A..19W}.

To further explore the above scenario, let us consider the synchrotron emission properties of the `Head' region which represents the innermost region in the eastern lobe characterized by the unusually hard photon index (of 1.58).
Assuming equipartition between energy in the relativistic electrons and magnetic fields, a volume of 1.58$\times$10$^{57}$~cm$^{-3}$,
the observed X-ray luminosity of 1.1$\times$10$^{34}$~ergs~s$^{-1}$ (in the 0.3--30 keV range) yields, in the synchrotron emission interpretation, an equipartition magnetic field value B of $\sim$12$\mu$G. This estimate should be considered a lower limit since the volume could be smaller, as suggested by the clumpy structures seen in the Chandra image.
The energy required to accelerate the electrons to X-ray energies in the `Head'
is approximately 2.4$\times$10$^{44}$~ergs.
This yields a radiative loss timescale of the order of 1~kyr.
We note that this is comparable to the synchrotron loss timescale (of $\sim$250--1,400 yr) of 1--30~keV photons in a 12$\mu$G magnetic field.
Therefore we find that the loss timescale is much smaller than the age of the system of $\lesssim$30~kyr
\citep{1997ApJ...483..868S, 1998AJ....116.1842D,2011MNRAS.414.2838G}. This suggests the presence of freshly ($\lesssim$kyr timescale) injected particles in the acceleration zone.

It is commonly believed that the SS~433 jets impact the W50 nebula out to the radio ears, with a significant fraction of the jets' power going into kinetic motion or an unseen outflow. 
In particular, the X-ray luminosity in the eastern lobe ($L_x$$\sim$10$^{35}$~erg~s$^{-1}$) 
is a very small fraction of the SS~433 jets' power. 
The latter ranges from 3.2$\times$10$^{38}$~erg~s$^{-1}$  based on Chandra HETG data \citep{Marshall2002}
 to 5$\times$10$^{39}$~erg~s$^{-1}$ based on XMM-Newton data \citep{2005A&A...431..575B}
 to $\sim$10$^{40}$~erg~s$^{-1}$ based on ASCA data \citep{2006ApJ...637..486K}.
Our study shows that the hard X-ray emission
starts at $\sim$29~pc from SS~433 with a gap in X-ray emission between SS~433 and the `Head'. Interestingly, compact X-ray knots were observed in the arcsecond-scale X-ray jets emitted by SS~433; their rapid variability (by a factor of 1.5--2.0 and on time-scales of days to as short as hours) points to `pumping' (from SS~433) or a sequence of shocks, revealing a fast, unseen outflow in the jet \citep{2005MNRAS.358..860M}. 
This outflow from SS~433 could provide the seed particles eventually accelerated at the onset `Head' region of the eastern lobe. 
The jet travel time between SS~433 and the `Head' is $\sim$370~yr assuming a distance of 29~pc and a constant velocity of 0.26c. However, it is expected that the jet outflow has significantly decelerated 
by the time it has travelled that far away from SS~433.
The synchrotron loss timescale of $\sim$1.5~kyr (for a 1 keV photon in a magnetic field of the order of 12~$\mu$G) is smaller than the travel time between the SS~433 jet and the `Head' if the jet's velocity is smaller than 0.065c (19,500~km~s$^{-1}$), a reasonable assumption. 
Therefore, it is likely that the hard X-ray emitting knots observed in the `Head' region are associated with the arcsecond-scale jet outflow and that the seed particles are accelerated through some non-traditional acceleration mechanism (as described above) or recently re-accelerated along the jet propagation.
This phenomenon is seen in extragalactic jets \citep{2003ASPC..300..151H}; at the acceleration site (where the X-rays are observed), the hard spectrum component of the synchrotron emission is expected to be visible only at the highest frequencies. This trend is also seen in PWNe where the onset of their X-ray emission corresponds to the pulsar wind termination shock (e.g., \cite{2008AIPC..983..171K}) and where a hard spectral index is observed to steepen away from the engine.

The detection of hard X-ray emission up to energies of $\sim$30~keV implies electron energies, E$_e$, given by (E$_e$/10 TeV)$\sim$0.5(B/1 mG)$^{-1/2}$(E$_{\gamma}$/1 keV)$^{1/2}$ (e.g., \cite{1995Natur.378..255K}). For an estimated field of 12$\mu$G, the corresponding electron energies are $\sim$250~TeV. This supports the scenario where this unique microquasar represents an efficient site for particle acceleration up to 100's of TeV energies.
Detailed SED modelling (for leptonic versus hadronic emission models, see e.g. \cite{2020ApJ...889..146S}) and interpretation in the light of interaction between the SS~433 jet and the SNR shell will be the subject of future work which will make use of our spatially resolved X-ray spectroscopy across the eastern lobe, combined with upcoming additional TeV studies from HAWC, HESS and hopefully LHAASO.

\section{Conclusions}
We presented the first joint NuSTAR and XMM-Newton observation of the eastern lobe, as well as the first
Chandra snapshot of its innermost region. 
Our results can be summarized as follows:
\begin{itemize}
    \item{We have detected, resolved and characterized the hard X-ray emission from the eastern lobe of W50. The eastern jet starts at $\lesssim$18$^{\prime}$ ($\sim$ 29~pc at $d_{\earth} = 5.5$~kpc) from SS~433 (the `Head') and is knotty with an $\sim$1$^{\prime}$ (1.6~pc) scale.}
    \item{The emission from the `Head' is characterized by a hard non-thermal X-ray spectrum with a photon index $\lesssim$1.6 (confirmed with NuSTAR, XMM-Newton and Chandra).}
    \item{The photon index gradually steepens away from SS~433 to $\sim$1.8 in the `Cone' and `Diffuse' regions, $\sim$2 in the `lenticular' and termination shock regions; with soft thermal X-ray emission ($kT$$\sim$0.2~keV) dominating the termination shock region at the radio `ear'.}
\end{itemize}

 The hard X-ray knots mark the location of acceleration sites within the jet;
the hard index challenges classical particle acceleration processes and suggests the need for an additional energizing of particles near the `Head' region.
Deep and high-resolution observations of all regions in W50 (underway or planned), combined with multi-wavelength SED modelling, 
will be the subject of future works aimed at completing our understanding of this unique system whose study is relevant to probing the outflows of many galactic and extragalactic sources, including PWNe, AGNs and ULX's.

\vspace{0.25cm}

\textit{Facilities:} NuSTAR, XMM-Newton, CXO

\textit{Software:}
HEAsoft (v6.28, HEASARC 2014), XMM-SAS (v.19.1.0, ESA XMM-Newton Science Analysis System 2020), CIAO (v4.12, \cite{2006SPIE.6270E..1VF}), NuSTARDAS (v2.0.0, HEASARC-NuSTAR and Caltech-NuSTAR Data Analysis pages), SAOImage~DS9 (\cite{2003ASPC..295..489J}, Smithsonian Astrophysical Observatory 2000), XSPEC v12.11.1 (\cite{1996ASPC..101...17A}).

\textit{Acknowledgments}
This research made use of NASA's Astrophysics Data System (ADS) and of data and software provided by the High Energy Astrophysics Science Archive Research Center (HEASARC), which is a service of the Astrophysics Science
Division at NASA/GSFC.
This work made use of data from the NuSTAR mission, a project led by the California Institute of Technology, managed by the Jet Propulsion Laboratory, and funded by NASA.
This work also made use of data obtained with XMM--Newton, a European Space Agency science mission with instruments
and contributions directly funded by ESA Member States and NASA. 
We thank Panos Boumis for providing the optical image, Gloria Dubner for the radio image, and Brenda Dingus for her contribution to the original observing proposal. 
We acknowledge discussions with Dmitry Khangulyan, Takahiro Sudoh and Naomi Tsuji related to the upcoming modelling and planning future observations of the source. We are grateful to the anonymous referee for a constructive report and very helpful comments that helped
improve the manuscript.
S.S.H. acknowledges support from the Natural Sciences and Engineering Research Council of Canada (NSERC) through the Discovery Grants and Canada Research Chairs programs, and from the Canadian Space Agency. B.M.I. is supported by a University of Manitoba Graduate Fellowship. K.F. acknowledges support from the Wisconsin Alumni Research Foundation, from NASA through the Fermi Guest Investigator Program (NNH19ZDA001N-FERMI, NNH20ZDA001N-FERMI) and from National Science Foundation (PHY-2110821). 



\appendix
\restartappendixnumbering

\section{Appendices:  Additional Material}

\subsection{Background analysis -- NuSTAR background simulations} \label{sec:nustarbkg}

In what follows, we 
detail the specifics for the NuSTAR background analysis given its complex nature and potential impact on the spectral results. 
A diffuse X-ray source such as the eastern lobe requires analyzing the NuSTAR background contamination carefully and optimizing the background extraction regions. In general, the NuSTAR background is composed of (1) focused (2-bounce) diffuse photons, (2) ghost-ray (1-bounce) photons from bright X-ray sources located within $\sim3^{\prime}-40^{\prime}$ off-axis angles, and (3) stray-light (0-bounce) photons from X-ray sources within $\sim1^\circ-5^\circ$ off-axis angles  \citep{Madsen2015}. The {\tt nuskybgd} software tool is not applicable since it does not take into account additional background contamination from the Galactic Ridge X-ray emission and ghost-rays from SS433. A more detailed investigation of the ghost-ray background analysis will be presented in our forthcoming paper on the western lobe where the background contamination is more severe due to its proximity to SS~433.

\begin{figure}
\begin{center} 
 \includegraphics[width=\textwidth]{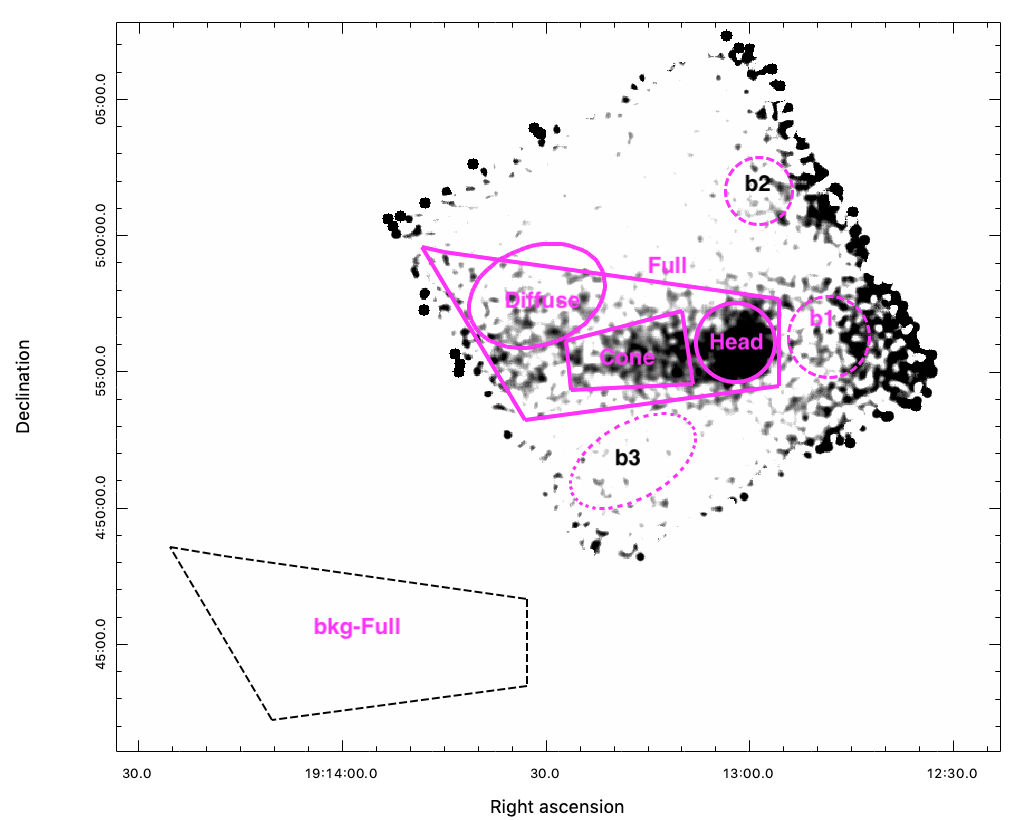}
  \caption{NuSTAR 3--10 keV image showing the background regions selected for our background analysis detailed in 
  \S\ref{sec:nustarbkg}.
 The solid purple regions are the source regions selected for our main spectral analysis, while the dashed regions represent the background regions (b1, b2, b3) explored for optimizing the background selection. For the `Full' source region, we chose a background from our off-source observation (dashed black polygon) with a similar PA. The final background regions selected for each of the source regions are shown in Figure~\ref{fig:regionsimages}.}
\end{center}
\label{fig:nustarbkgima}
\end{figure}

The off-source NuSTAR observation provides the most accurate stray-light background count rate maps since it was performed at a similar PA as
the on-source observation  \citep{Mori2015}. 
We compared the count rates from three source-free regions (b1, b2 and b3, as indicated in the NuSTAR image 
(Figure~\ref{fig:nustarbkgima})
 in the FPMA and FPMB images to the corresponding regions in the off-source observation images. In the region to the west of the `Head' (b1), the count rates for both modules are higher than those extracted from the off-source observation. Our NuSTAR imaging analysis in  \S\ref{sec:imaging}
revealed the ghost-ray contamination clearly in the spatial profiles produced in different energy bands. We also found that background spectra extracted from b1 and b2 showed a hint of Fe emission lines at $E\sim$6--7 keV which are not expected from the non-thermal X-ray emission in the eastern lobe. 
We attributed the count rate discrepancy and Fe lines to ghost-ray background photons from SS~433. 
In the b3 region, we found a good agreement in the count rates between the on-source and off-source observations, suggesting that it's not impacted by the ghost rays (as for b1 and b2) and that the stray-light background component is predominant. 

We further investigated the level of background contamination using the off-source NuSTAR observations and NuSIM simulator. NuSIM is a Monte Carlo ray-tracing simulation tool designed for the NuSTAR mission \citep{Zoglauer2011}. While the off-source observations are used for estimating stray-light background count rates, NuSIM allows us to simulate ghost-ray events from SS~433. 
We simulated ghost-ray background events from SS~433 by inputting an X-ray flux of $3.7\times10^{-3}$ photons\,s$^{-1}$\,cm$^{-2}$ (3--79 keV) which corresponds to the Swift-BAT X-ray flux averaged over $\pm1$ days around the NuSTAR observation of the eastern lobe. We assumed a power-law model of $\Gamma = 1.35$  as a typical X-ray spectrum of SS~433 \citep{Marshall2002}. We generated $\sim10^5$ ghost-ray background events to reduce statistical errors and visualize their spatial distributions clearly. 
Then, we estimated the NuSTAR background count rates in the three source-free regions by adding the actual stray-light and simulated ghost-ray background count rates. We found that the observed and estimated background count rates in the three regions are consistent with each other within statistical errors and the Swift-BAT flux variation during the NuSTAR observation. In the b1 and b2 regions which are closer to SS~433, the ghost-ray component is more dominant accounting for $\sim50$\% of the total background count rates. Therefore, we extracted background spectra from b1 for the `Head' region and found consistent spectral fitting results with the XMM-Newton spectra in the 3--10 keV band (Table~\ref{tab:pow0.5-30}). The NuSIM simulation maps showed that the ghost-ray count rates decrease radially with the angular distance from SS~433. Beyond the b1 and b2 regions, the ghost-ray contribution is negligible as is also evident from the fact that the count rates in b3 match with those from the (stray-light dominating) off-source observations. 

For the `Cone' and `Diffuse' regions where the ghost rays background should be minimal but the stray-light background component should be most significant, we attempted background spectra from the same detector regions (as the source extraction regions) in the off-source observation, and compared the spectral fitting using local, nearby source backgrounds. We do not find any significant difference and so
we opted for the local background regions (shown in Figure~2)
as for the XMM-Newton region selection, especially as this background removes also any contamination from the Galactic ridge X-ray emission.
Finally for the larger `Full' region, which encompasses all NuSTAR source regions, we selected an off-source background (shown as the dashed black polygon south-east of the source FoV) primarily because of the region's large size and our limited choice for a local background. We believe that the apparent discrepancy between the XMM-Newton and NuSTAR spectral result for this region in particular (Table~\ref{tab:pow0.5-30})
is partly attributed to the impact of the NuSTAR background subtraction, but mostly because this region consists of regions characterized by different spectral components, as shown in \S\ref{sec:spresspec}.

\newpage

\begin{table}[H]
    \centering
    \begin{tabular}{c|c|c|c|c|c}
    \hline
   Satellite & Date & ObsID & PI & Pointing & Total Exposure \\
    & & & & RA, Dec (J2000) & (ks) \\
    \hline
        NuSTAR  & 2019-12-01 & 40510002001  & Safi-Harb, S. & 19 13 11.0,	+04 57 03 & 109,419 \\
          &  & 40510003001 &  & 19 13 47.1,	+04 46 21 (background) & 32,358 \\
        XMM-Newton & 2020-03-24 & 0840490101 & Safi-Harb, S. & 19 13 13.37,	+04 57 29.8 & 69,100  \\
        XMM-Newton & 2004-09-30 & 0075140401 &  Brinkmann, W. & 19 14 12.01,	+04 55 47.0	 & 32,513 \\
        XMM-Newton & 2004-10-04 & 0075140501 & Brinkmann, W. & 19 15 55.01,	+04 51 20.0	& 31,314\\
        Chandra & 2000-06-27 & 659 & Fender, R. & 19 11 49.50, +04 58 58.0 & 9,790 \\
        \hline
    \end{tabular}
    \caption{Log of the X-ray observations presented in this study}
    \label{tab:observationslog}
\end{table}

\begin{table}[H]
    \centering
    \begin{tabular}{llccc}
                                                                                 & Parameter    & XMM 0.5 -- 10 keV & NuSTAR 3 -- 30 keV & Joint 0.5 -- 30 keV$^c$ \\ \hline
    \multirow{4}{*}{\begin{tabular}[c]{@{}l@{}}Head\\ $N_{\rm H}$ = 1.77$\times$10$^{22}$~cm$^{-2}$\end{tabular}}    & Photon Index $\Gamma$ & $1.58 \pm 0.06$   & 1.6 $\pm$ 0.1   & $1.58 \pm 0.05$      \\
                                                                                 &  $F\left[\times 10^{-12}\right]^{a,c}$ (abs.)      & $1.23 \pm 0.06 $& $2.0 \pm 0.1$  &   $2.45 \pm 0.07$ \\

                                                                                 &
                                                  $F\left[\times 10^{-12}\right]^{b,c}$ (unabs.) &       $1.80 \pm 0.06$ &   $2.0 \pm 0.1$  &       3.0 $\pm$ 0.1 \\                                                                                             &
                                          $\chi^2_\nu$~(DoF)          & $0.96 (295)$   & $1.19 (79)$   & $1.00~(375)$      \\ \hline
    \multirow{4}{*}{\begin{tabular}[c]{@{}l@{}}Cone\\ $N_{\rm H}$ = 1.18$\times$10$^{22}$~cm$^{-2}$\end{tabular}}    & Photon Index $\Gamma$ & $1.65 \pm 0.05$   & $2.00_{-0.07}^{+0.08}$   & $1.76 \pm 0.04$      \\
                                                                                 &
                                                                                 $F\left[\times 10^{-12}\right]^a$ (abs.)     &  $1.29\pm 0.05$ & $1.55 \pm 0.09$ & $2.13 \pm 0.06$ \\ 
                                                                                                                                                              &
                                                $F\left[\times 10^{-12}\right]^b$ (unabs.) &   $1.81 \pm 0.06$ &   $1.58_{-0.09}^{+0.10}$ & $2.71 \pm 0.08$ \\
                                                                                  & $\chi^2_\nu$~(DoF)          & $0.96~(440)$   & $0.98~(197)$   & $1.03~(638)$      \\ \hline
    \multirow{4}{*}{\begin{tabular}[c]{@{}l@{}}Diffuse\\ $N_{\rm H}$ = 0.76$\times$10$^{22}$~cm$^{-2}$\end{tabular}} & Photon Index $\Gamma$ & $1.75 \pm 0.06$   & $2.0 \pm 0.2$   & $1.77 \pm 0.06$      \\
                                                                                 &
                                                  $F\left[\times 10^{-12}\right]^a$ (abs.) &    $1.02_{-0.05}^{+0.06}$ & $1.00_{-0.14}^{+0.05}$ &   $1.61 \pm 0.08$ \\
                                         
                                                                                 &
                                                 $F\left[\times 10^{-12}\right]^b$ (unabs.) &  $1.39 \pm 0.06$ &    $1.00_{-0.14}^{+0.15}$ & $1.99 \pm 0.09$ \\
                                                                                                                 & $\chi^2_\nu$~(DoF)          & $0.98~(467)$   & $1.33~(59)$   & $1.02~(527)$      \\ \hline
   \hline                                                                              
    \multirow{4}{*}{\begin{tabular}[c]{@{}l@{}}Full\\ $N_{\rm H}$ = 1.03$\times$10$^{22}$~cm$^{-2}$\end{tabular}}    & Photon Index $\Gamma$ & $1.58 \pm 0.03$   & $1.99 \pm 0.07$   & $1.65 \pm 0.03$      \\
                                             
                                                                                 &
                                                                                 $F\left[\times 10^{-12}\right]^a$ (abs.)      & 5.60 $\pm$ 0.14 & $8.8 \pm 0.4$ & $11.4 \pm 0.2$ \\  
                                                                                 &
                                                     $F\left[\times 10^{-12}\right]^b$ (unabs.) & $7.5 \pm 0.2$ & $9.0 \pm 0.4$ & 13.4 $\pm$ 0.2 \\

                                                                                  & $\chi^2_\nu$~(DoF)          & $1.03~ (1589)$   & $0.97~(98)$   & $1.07~(1688)$      \\ \hline
    \multirow{4}{*}{\begin{tabular}[c]{@{}l@{}}Nue1\\ $N_{\rm H}$ = 0.78$\times$10$^{22}$~cm$^{-2}$\end{tabular}}    & Photon Index $\Gamma$ & $1.73 \pm 0.05$   & $2.2 \pm 0.2$   & $1.76 \pm 0.05$      \\
                                                                                 &
                                                    $F\left[\times 10^{-12}\right]^a$ (abs.)
                                                    & $1.62 \pm 0.07$ & $2.0 \pm 0.2$ & $2.89 \pm 0.19$ \\
                                                                                  &
                                                                                 
                                            $F\left[\times 10^{-12}\right]^b$ (unabs.) &     $2.20 \pm 0.08$ &  $2.0 \pm 0.2$ & $3.48 \pm 0.12$ \\
                                                                              & $\chi^2_\nu$~(DoF)          & $1.05~(652)$   & $1.53~(35)$   & $1.09~(688)$      \\ \hline                                                                        \hline
    \multirow{4}{*}{\begin{tabular}[c]{@{}l@{}}Lenticular$^d$
    \\ $N_{\rm H}$ = 0.71$\times$10$^{22}$~cm$^{-2}$\end{tabular}}    & Photon Index $\Gamma$ & $2.05 \pm 0.02$   & --   & --      \\
                                                                                 & $F\left[\times 10^{-12}\right]^a$ (abs.)      & $3.74 \pm 0.06$   & --   & --      \\
                                                                                 & $F\left[\times 10^{-12}\right]^b$ (unabs.)        & $5.68 \pm 0.08$   & --   & --      \\
                                                                                 & $\chi^2_\nu$~(DoF)          & $1.04~(1101)$   & --   & --      \\ \hline
                                                     
    \multirow{4}{*}{\begin{tabular}[c]{@{}l@{}}e3-ring
    \\ $N_{\rm H}$ = 0.79$\times$10$^{22}$~cm$^{-2}$\end{tabular}}    & Photon Index $\Gamma$ & $2.0^{+0.4}_{-0.5}$  & --   & --      \\
                                                                                              & $kT$ (keV)      & 0.2 $\pm$ 0.1   & --   & --      \\                 
                                                                                 & $\chi^2_\nu$~(DoF)          & 1.14~(797)   & --   & --      \\ \hline
    \end{tabular}
    \caption{XSPEC fits using {\tt const*tbabs*power} with the {\tt cFlux} model added to acquire flux values. Column densities frozen to their best fit value using the XMM-Newton data. For the `e3-ring' region, the additional component shown corresponds to a thermal ({\tt mekal} in XSPEC) model with solar abundances. Quoted uncertainties for 90\% C.L. \\
    $^a$Absorbed flux in erg cm$^{-2}$ s$^{-1}$.\\
    $^b$Unabsorbed flux in erg cm$^{-2}$ s$^{-1}$. \\
    $^c$ For the joint 0.5--30 keV flux, we list the combined 0.5--10 keV (XMM) and 10-30 keV (NuSTAR) fluxes with the model parameters frozen to their best values from the joint fit.\\ 
    $^d$ The larger `e2' region encompassing the `lenticular' region (see 
Figure~\ref{fig:regionsimages}) shows evidence of soft thermal X-ray emission with $kT$$\lesssim$0.2~keV, however with the thermal parameters poorly constrained.\\
}
    \label{tab:pow0.5-30}
\end{table}

\pagebreak

\bibliographystyle{aasjournal}
\bibliography{ms}

\end{document}